\documentclass[useAMS,usenatbib]{mnras}
\usepackage{graphicx}
\usepackage{amsmath}
\usepackage{amssymb}
\usepackage{comment}
\usepackage[normalem]{ulem}

\include{defns}
\def\gsim{\;\rlap{\lower 2.5pt
 \hbox{$\sim$}}\raise 1.5pt\hbox{$>$}\;}
\def\lsim{\;\rlap{\lower 2.5pt
   \hbox{$\sim$}}\raise 1.5pt\hbox{$<$}\;}
\defcitealias{liu09a}{Liu et al. (2009a)}
\defcitealias{liu09b}{Liu et al (2009b)}

\newcommand{\tr}[1]{\textrm{#1}}
\newcommand{\ee}[1]{\times10^{#1}}

\newcommand{\pp}[2]{\frac{\partial#1}{\partial#2}}
\newcommand{\mbfB}{\mathbf{B}}

\newcommand{\kpar}{k_{\parallel}}
\newcommand{\mbfkpar}{\mathbf{k_{\parallel}}}

\title[]{High Beta Effects on Cosmic Ray Streaming in Galaxy Clusters}
\author[Wiener, Zweibel, \& Oh]{Joshua Wiener$^{1,2}$, Ellen G. Zweibel$^{2,3}$, \& S. Peng Oh$^{1}$\\
$^{1}$ Department of Physics; University of California; Santa Barbara, CA 93106, USA.\\
$^{2}$ Department of Astronomy, University of Wisconsin-Madison, Madison, WI 53706, USA.\\
$^{3}$ Department of Physics, University of Wisconsin-Madison, Madison, WI 53706, USA.}

\begin{document}
\bibliographystyle{mnras}

\pagerange{000--000} \pubyear{0000}
\maketitle

\label{firstpage}

\begin{abstract} Diffuse, extended radio emission in galaxy clusters, commonly referred to as radio halos, indicate the presence of high energy cosmic ray (CR) electrons and cluster-wide magnetic fields. We can predict from theory the expected surface brightness of a radio halo, given magnetic field and CR density profiles. Previous studies have shown that the nature of CR transport can radically effect the expected radio halo emission from clusters \citep{wiener13a}. Reasonable levels of magnetohydrodynamic (MHD) wave damping can lead to significant CR streaming speeds. But a careful treatment of MHD waves in a high $\beta$ plasma, as expected in cluster environments, reveals damping rates may be enhanced by a factor of $\beta^{1/2}$. This leads to faster CR streaming and lower surface brightnesses than without this effect. In this work we re-examine the simplified, 1D Coma cluster simulations (with radial magnetic fields) of \cite{wiener13a} and discuss observable consequences of this high $\beta$ damping. Future work is required to study this effect in more realistic simulations.
\end{abstract}

\section{Introduction}
\label{sec:intro}
Galaxy clusters are the largest gravitationally bound objects in the universe, and are host to processes described by every area of physics. For some clusters, this includes particle acceleration to very high energies, generating cosmic rays (CRs). Cosmic ray electrons (CRe) at these energies will emit synchrotron emission in the presence of magnetic fields which is observable as radio emission.

Galaxy clusters can exhibit radio emission of a few morphological types. We focus here on radio haloes, which refers to extended, diffuse radio emission. Studies have shown that where radio haloes are detected, the total radio luminosity of the halo correlates well with total X-ray luminosity of the host cluster. But many clusters do not have radio haloes at all. Thus there is a peculiar bimodality in the presence of radio haloes in galaxy clusters that must be explained (\cite{brunetti07b}, \cite{ensslin11}, \cite{basu12, sommer14}).

Related to this problem is the source of the CRe which produce the emission. CRe at the required energies ($\sim$1-10 GeV) have short cooling times ($\lsim$100 Myr) for typical cluster magnetic field strengths ($\sim$3 $\mu$G) and densities ($\sim 10^{-3}\ \tr{cm}^{-3}$). The presence of a radio halo in a given cluster therefore indicates relatively recent injection of CRe. The `hadronic model' claims that this is provided by hadronic collisions of CR protons (CRp) with ambient thermal nuclei that result in pions, which then decay into secondary CRe. A rival model, `turbulent reacceleration', hypothesizes that low-energy CRe undergo Fermi-II reacceleration by gas motions during mergers \citep{brunetti07}. However, the abundance and distribution of CRp is still an important ingredient in this model, as secondary CRe produced during hadronic collisions provide seeds for reacceleration \citep{brunetti11,pinzke17}.\footnote{Acceleration from the thermal pool is precluded by strong Coulomb losses \citep{petrosian08}.} An excellent summary of the state of the field can be found in \citet{brunetti14}.

An issue with the hadronic model is that high energy CRp have much longer ($\sim$10-100 Gyr) cooling times. If a cluster undergoes some merger or structure formation process that may be responsible for accelerating CRs, then we may naively expect the resulting radio halo to last for as long as or longer than the age of the universe. If this were the case, we would struggle to explain clusters that don't exhibit radio halo emission, since every cluster is expected to accelerate CRp at some point in its history through such a mechanism. A lack of hadronic emission implies either that the assumed acceleration efficiencies are too high or that the CRp are quickly diluted by transport effects. We investigate this second possibility. Note that CRp are frequently invoked in turbulent reacceleration models as well, to generate secondary seed electrons for reacceleration. Observations are best fit by a flat CRp profile, which can be explained by CR streaming \citep{pinzke17}.

Building on the work of \cite{ensslin11}, \cite{wiener13a} attempted to resolve this issue by suggesting that, for favorable magnetic field orientations, CRp can travel at high speeds away from cluster centers, quickly reducing the CRp density and thus the expected radio luminosity on appropriate time scales. Bulk CR transport is limited by the streaming instability, which amplifies hydromagnetic waves traveling in the same direction as the mean cosmic ray velocity if the cosmic rays are anisotropic in the frame of the wave \citep{kulsrud69}. However, if other plasma processes damp the wave then a larger streaming speed is required to excite the instability, resulting in faster cosmic ray transport. In \cite{wiener13a} it was shown that under galaxy cluster plasma conditions the strongest damping mechanism is the turbulent damping process proposed by \cite{yan02,farmer04}. 

Galaxy cluster plasmas are characterized by large ratios of thermal to magnetic pressure, a parameter usually denoted as $\beta$. In this work we show that, in high $\beta$ environments, wave damping is actually stronger than the rate quoted in \cite{wiener13a}. Therefore, CRs can travel even faster than previously thought and radio halos turn off faster than we previously estimated for the same density and magnetic field profile\footnote{As we discuss in \S\ref{sec:topology}, we use a very simple 1D halo model with purely radial magnetic fields for comparison with \cite{wiener13a}. More sophisticated simulations are needed in future work to examine the effects of this damping in real clusters.}. While this may explain the bimodality of radio haloes in the hadronic model, it cannot reconcile other issues with the hadronic model. Among others, gamma-ray non-detections in the Coma cluster put strict limits on the amount of CRp present (\cite{brunetti12, zandanel14b, fermi16, brunetti17}). As such, this work should not be seen as an attempt to strengthen the hadronic model of radio haloes so much as a more accurate theory of CR transport in clusters in general, which is still an important ingredient in the reacceleration model (\cite{pinzke17}).

Also, although the time evolution of radio halos is the main focus of the paper, we note that faster streaming may also help to explain the puzzling lack of diffuse $\gamma$-rays from galaxy cluster cores (\citet{pfrommer08,pinzke10, ensslin11,ahnen16}). It may also be germane to understanding radio mini-halo luminosities \citep{jacob17}. While the hadronic model is problematic for understanding giant radio halos, it remains the chief contender for understanding kinematically quiescent radio mini-halos \citep{zandanel14}, where the CRp could be sourced by a central AGN. Note that modulo loss processes, the contribution of the AGN jet to CRe could be non-negligible as well.

In \S\ref{sec:damping} we review the theory of MHD wave damping in collisionless, high $\beta$ plasmas and show how it modifies previously estimated turbulent damping rates. An important part of this calculation is an estimate of the effective collisionality of the plasma under the assumption that microturbulence is present due to pressure anisotropies \citep{schekochihin07}; we argue that the level of microturbulence is low enough that Alfv\'en waves at the characteristic wavelengths excited by cosmic ray steaming instabilities are nevertheless collisionlessly damped. We also discuss the possible effects of the magnetic field topology, which is not taken into account in our simulations. In \S\ref{sec:sim} we describe the simulation setup, which is similar to the setup we used previously for the Coma cluster \citep{wiener13a}. Sections \S\ref{sec:results} and \S\ref{sec:conclusion} gives the results and conclusion, respectively.

\section{MHD Waves in Turbulent, High Beta Plasma}\label{sec:damping}

\subsection{Wave damping}
As mentioned in \S\ref{sec:intro}, the streaming speed as a function of energy is determined by balancing the growth rate of the streaming instability \citep{kulsrud71}
\begin{equation}
\label{Gammacr}
\Gamma_{cr}(\mbfkpar) = \omega_{cp}\frac{n_{CR}(>\gamma_R)}{n_i}\left(\frac{v_D(\gamma_R)}{v_A}-1\right)
\end{equation}
against the mechanism(s) that damp the wave. In (\ref{Gammacr}), $\omega_{cp}$ is the proton gyrofrequency (for proton cosmic rays), $\gamma_R\equiv\omega_{cp}/c\kpar$ is the minimum Lorentz factor of a resonant cosmic ray, and $v_D$ is the energy dependent streaming speed.

Equation (\ref{Gammacr}) is computed assuming a perfectly straight and uniform background magnetic field $\mbfB_0$. Under this assumption, the fastest growing waves propagate parallel to $\mbfB_0$, with waves of any given wavenumber $\kpar$ being driven primarily by cosmic rays with gyroradii $r_L\sim\kpar^{-1}$.

If $\mbfB_0$ has curvature or perpendicular structure,
strict parallel propagation is impossible: there is a minimum angle $\theta_{min}$ which depends on the background field structure. \cite{farmer04} evaluated
$\theta_{min}$ for Alfv\'en waves excited by cosmic ray streaming in a background field upon which an anisotropic MHD turbulent cascade is imposed, and found it to be
\begin{equation}\label{eq:minAngle2}
 \qquad \tan\theta_{min} =
\left(\frac{\lambda_\parallel}{\lambda_\perp}\right) \sim
\left(\frac{r_L}{L_\tr{MHD}}\right)^{1/4} \sim 
\left(\frac{r_L\epsilon}{v_A^3}\right)^{1/4},
\end{equation}
where $L_\tr{MHD}$ is the length scale at which the characteristic bulk speed is the Alfv\'en speed $v_A$, and the turbulent energy cascade rate is $\epsilon\equiv v_A^3/L_\tr{MHD}$. The perpendicular lengthscale $\lambda_{\perp}$ and turbulent velocity $v_{\lambda_\perp}$ corresponding to $\theta_{min}$ are
\begin{equation}\label{eq:minAngle}
 \qquad \lambda_\perp \sim r_L^{3/4} L_\tr{MHD}^{1/4} \sim r_L^{3/4}(v_A^3/\epsilon)^{1/4},
\end{equation}
and 
\begin{equation}
\label{vlambdaperp}
v_{\lambda_\perp}\sim v_A\left(\frac{\lambda_{\perp}}{L_\tr{MHD}}\right)^{1/3}\sim v_A\left(\frac{r_L}{L_\tr{MHD}}\right)^{1/4}.
\end{equation}

According to the \cite{farmer04} scenario, MHD waves are progressively sheared by the turbulent cascade, eventually transferring their power to the dissipation scale. The corresponding ``turbulent damping rate" is therefore on the order of the eddy turnover rate at the perpendicular scale
\begin{equation}
\label{Gammaturb}
\Gamma_\tr{damp} \sim \frac{v_{\lambda_\perp}}{\lambda_\perp} \sim \frac{\epsilon^{1/3}}{\lambda_\perp^{2/3}}\sim \left(\frac{\epsilon}{r_Lv_A}\right)^{1/2},
\end{equation}
where in the last equality we have used eqns. (\ref{eq:minAngle}) and
(\ref{vlambdaperp}). This was the damping rate used in the \small{ZEUS} simulations in \cite{wiener13a}. The \citet{farmer04} calculations have recently been generalized by \citet{lazarian16} to a variety of scenarios.

A key assumption in deriving eqn. (\ref{Gammaturb}) is that the dissipation scale is much shorter than the wavelength of the wave. If this is not the case, the dissipation rate can exceed the shearing rate. 

In collisionless, high $\beta$ plasmas, MHD waves of even small obliquity are subject to strong ion Landau damping \citep{foote79}, a process whereby ions with parallel velocity $v_{\parallel}$ which satisfies the resonance condition $v_{\parallel}=\omega/\kpar$ absorb energy from the wave due to acceleration by its parallel electric field {\footnote{This process is not to be confused with {\textit{nonlinear}} Landau damping, in which the thermal ions resonate with the low frequency beat wave generated by the interaction of two higher frequency waves \citep{kulsrud05}. NLLD was first invoked for galaxy cluster plasmas by \cite{loewenstein91} and was argued to be subdominant in turbulent galaxy cluster plasmas in \cite{wiener13a}.}}. 

We briefly review the Foote \& Kulsrud calculation here. They defined
\begin{equation}
\beta = \frac{P_g}{P_B} = \frac{\rho k_BT/\mu m_p}{B^2/8\pi} = \frac{v_i^2}{v_A^2},
\end{equation}
and framed their analysis in terms of the rescaled wave frequency $\omega$ and wave number $k$:
\[
\nu \equiv \frac{\omega}{k_0v_A}, \quad l \equiv \frac{k\cos\theta}{k_0}, \quad k_0 \equiv \frac{\omega_{ci}v_A}{v_i^2}
,\]
where $\omega_{ci}$ is the thermal ion cyclotron frequency. They also introduce a parameter $\alpha=\pi^{1/2}\beta^{1/2}l^2\tan^2\theta$.

\cite{foote79} show that in the small $l$ limit, which applies to most of the waves amplified by streaming cosmic rays, the dispersion relation can be approximated asymptotically by
\begin{equation}
\label{nu}
\nu = l - \frac{i\alpha}{4l} \pm \frac{1}{4l}\left(l^6-\alpha^2+\frac{4i\alpha^3}{l^2}\right)^{1/2}
\end{equation}
Expanding eqn. (\ref{nu}) for small $\alpha$, the damping rate is given by
\begin{equation}
\Gamma_\tr{damp} = -k_0v_A\tr{Im}(\nu) \approx k_0v_A\frac{\alpha}{4l}
\end{equation}
Using \eqref{eq:minAngle2} for $\theta$ and restoring dimensional units, we arrive at the result
\[
\Gamma_\tr{damp} \approx \frac{\sqrt{\pi}}{4}k_0v_A\beta^{1/2}l\tan^2\theta \approx \frac{\sqrt{\pi}}{4} \beta^{1/2}r_L^{-1}v_A\left(\frac{r_L\epsilon}{v_A^3}\right)^{1/2}
\]
\begin{equation}\label{eq:betadamp}
\Rightarrow \quad \Gamma_\tr{damp} \approx \frac{\sqrt{\pi}}{4} \beta^{1/2}\left(\frac{\epsilon}{r_Lv_A}\right)^{1/2},
\end{equation}
where we have used $lk_0 = k\cos\theta = r_L^{-1}$.

Comparison with \eqref{Gammaturb} shows that with this treatment the wave damping rate is enhanced by a factor of $\beta^{1/2}$. In the cluster environments simulated in \cite{wiener13a}, where $\beta$ can be of the order of 100 in the cluster centers, this factor can lead to significantly higher CR streaming speeds than those predicted by \eqref{Gammaturb}.

The factor of $\beta^{1/2}$ can be heuristically understood as follows. The Landau damping rate is given by the rate at which resonant ions absorb energy from oblique waves; $\Gamma_{\rm damp}^{\rm Landau} \sim k v_{i} {\rm tan}^{2} \theta$. The turbulent damping rate is given by the rate at which a pair of interacting Alfv\'en waves cascade, $\Gamma_{\rm damp}^{\rm turb} \sim k v_{\rm A} {\rm tan}^{2} \theta$. The geometrical factor of ${\rm tan}^{2} \theta$ is the same for these pairwise interactions. Thus, the Landau damping rate is larger by a factor $\sim v_{i}/v_{\rm A} \sim \beta^{1/2}$.

\subsection{Collisionality}
\label{subsec:collisionality}
The analysis by \cite{foote79} assumes a collisionless plasma. We must verify that the plasmas in galaxy cluster environments we will be simulating are sufficiently collisionless. In particular, we must determine whether microinstabilities driven by pressure anisotropy can render the thermal ions essentially collisional. The relevant comparison to make here is the ion mean free path to the typical wavelength of the Alfv\'en waves in question.

The long mean free paths,  relatively
weak magnetic fields, and pervasive large scale turbulence in galaxy cluster plasmas make them attractive candidates for shear driven pressure anisotropy \citep{schekochihin06} through distortion of the magnetic field and preservation of the particles' adiabatic invariants. We
follow the notation of \cite{kunz11} in the following discussion.

Shear in a fluid with velocity field $\mathbf{u}$ and magnetic field direction $\mathbf{b}$ will drive pressure anisotropy while collisions will oppose it. Balancing the two yields an equilibrium anisotropy given by
\cite{braginskii65} 
\begin{equation}
\Delta_i\equiv\frac{P_{\perp,i}-P_{\parallel,i}}{P_i}=
\frac{2.9}{\nu_{ii}}\left(\mathbf{bb}:\nabla\mathbf{u}-\frac{1}{3}\nabla\cdot\mathbf{u}\right).
\end{equation}
In the above, the $i$ subscript indicates we are referring to ions and $\nu_{ii}$ indicates the frequency of Coulomb collisions. These collisions serve to isotropize the pressure. 

A complication arises if collisions are so infrequent that the pressure anisotropy exceeds the threshold for microinstabilities:
\[
\Delta_i>\frac{1}{\beta_i},\quad\tr{mirror instability}
\]
\[
\Delta_i<-\frac{2}{\beta_i},\quad\tr{firehose instability}
\]
If Coulomb collisions alone are not sufficient to prevent these instabilities, we may expect two possible responses of the fluid. In one scenario, the shear $S\equiv\mathbf{bb}:\nabla\mathbf{u}-1/3\nabla\cdot\mathbf{u}$ will adjust itself to reduce the pressure anisotropy until marginal stability is achieved. In this case the collision frequency does not change - it is simply the Coulomb collision frequency and the mean free path is the Coulomb mean free path. This is the working hypothesis in \cite{kunz11}.

In the other scenario, the shear $S$ remains unchanged and instead the resulting magnetic fluctuations driven by the instabilities will themselves scatter the ions. This increases the effective total scattering frequency $\nu_{i}$, which now includes these magnetic scatterings in addition to Coulomb scattering, until marginal stability is achieved:
\begin{equation}\label{eq:marginal_stability1}
\Delta_i=\frac{2.9}{\nu_i}S=\frac{2\xi}{\beta_i} \quad \rightarrow \quad \nu_i=\frac{1.45\beta_i}{\xi}S,
\end{equation}
where $\xi$ is -1 for the firehose instability, and 1/2 for the mirror instability. In this case, the total collision frequency is enhanced, and so the mean free path is shorter than the Coulomb mean free path.

The above marginal stability criterion therefore provides a lower limit on the ion scattering frequency $\nu_i$, and thus an upper limit on the ion mean free path $\lambda_i=v_i/\nu_i$. This limit depends on the shear forcing $S$. 

Let us approximate this forcing as $S\sim U/L$ for some characteristic speed $U$ and length scale $L$.
If we assume a turbulent cascade with Kolmogorov scaling, $U^3/L=\tr{const.}$, we can relate any $U$ and $L$ to the outer scales $U_0$ and $L_0$. The marginal stability criterion then becomes
\begin{equation}\label{eq:marginal_stability2}
\nu_i\sim\frac{1.45\beta_i}{\xi}\frac{U}{L}=\frac{1.45\beta_i}{\xi}\frac{U_0}{L_0}\left(\frac{L_0}{L}\right)^{2/3}
\end{equation}
The highest collision frequency will therefore be dictated by the smallest scale of turbulence, the dissipation scale $L_d$. This scale is determined by balancing the cascade rate $U_d/L_d$ with the viscous dissipation rate, which itself depends on the collision frequency, and is of order $v_i^2/(\nu_iL_d^2)$. We obtain
\[
\frac{U_d}{L_d}=\frac{U_0}{L_0}\left(\frac{L_0}{L_d}\right)^{2/3}\sim\frac{v_i^2}{\nu_iL_d^2}
\]
\[
\Rightarrow\quad\left(\frac{L_d}{L_0}\right)^{4/3}\sim\frac{v_i^2}{\nu_iU_0L_0}
\]
\begin{equation}\label{eq:dissipation_length1}
\frac{L_d}{L_0}=\left(\frac{v_i^2}{\nu_iU_0L_0}\right)^{3/4}
\end{equation}
Combining \eqref{eq:marginal_stability2} and \eqref{eq:dissipation_length1} we arrive at an expression for the ion collision rate at marginal stability
\[
\nu_i\sim\frac{1.45\beta_i}{\xi}\frac{U_0}{L_0}\left(\frac{\nu_iU_0L_0}{v_i^2}\right)^{1/2}
\]
\begin{equation}\label{eq:collisionality}
\Rightarrow\quad\nu_i\sim\left(\frac{1.45\beta_i}{\xi}\right)^2\frac{U_0^3}{L_0v_i^2}
\end{equation}
indicating an ion mean free path of
\begin{equation}\label{eq:mfp}
\lambda_i\sim 0.48\xi^2\frac{L_0}{\beta_i^2}\left(\frac{v_i}{U_0}\right)^3\sim 0.48\frac{L_0}{\beta_i^{2}\mathcal{M}_0^{3}}
\end{equation}
where we have defined the Mach number at the driving scale $\mathcal{M}_0\equiv U_0/v_i$.

We can plug this solution for $\nu_i$ back in to \eqref{eq:dissipation_length1} to find the dissipation length:
\begin{equation}\label{eq:dissipation_length2}
L_d\sim L_0\left(\frac{v_i^2}{\beta_i^2U_0^4/v_i^2}\right)^{3/4}\sim L_0\mathcal{M}_0^{-3}\beta_i^{-3/2}
\end{equation}
Note that the turbulent velocity at this scale is
\[
U_d=U_0\left(\frac{L_d}{L_0}\right)^{1/3}\sim U_0\mathcal{M}_0^{-1}\beta_i^{-1/2}=v_A.
\]
That is, under the assumption of marginal stability, the characteristic turbulent velocity at the dissipation scale is about the same as the Alfv\'en speed. This means the dissipation scale $L_d$ is about the same as the length scale $L_\tr{MHD}$ defined in \cite{wiener13a}'s streaming simulations as a measure of the wave damping rate, where it is assumed to be of order 100 kpc.

Now we can make an estimate of the mean free path as a function of $L_\tr{MHD}$:
\begin{equation}\label{eq:marginal_mfp}
\lambda_i\sim L_0\beta_i^{-2}\mathcal{M}_0^{-3}\sim L_d\beta_i^{-1/2} \sim 14\ \tr{kpc}\left(\frac{L_\tr{MHD}}{100\ \tr{kpc}}\right)\left(\frac{\beta_i}{50}\right)^{-1/2}
\end{equation}
This is much longer than the several AU gyroradii of the $\sim$100 GeV CR protons that generate the secondary $e^{\pm}$ which produce the observed radio emission. The assumption of collisionless wave damping is therefore sound if we assume marginal stability to MHD microinstabilities.

We can also compare the above result to the Coulomb mean free path $\lambda_{ii}$. If $\lambda_{ii}$ is significantly shorter, then Coulomb collisions alone are enough to keep the pressure anisotropy \eqref{eq:marginal_stability1} low enough to avoid MHD microinstabilities. We have
\begin{equation}
\label{eq:coulomb_mfp}
\lambda_{ii}=v_it_i=\sqrt{\frac{kT}{m_p}}\frac{\sqrt{m_p}(kT)^{3/2}}{4\sqrt{\pi}\ln\Lambda e^4 n_i}\approx 5\ee{20}\ \tr{cm}\frac{T_7^2}{n_{i,-3}}
\end{equation}
where we approximate $\ln\Lambda\approx 30$ and we adopt the subscript notation $Q_x=Q/10^x$ in cgs units.

In the central regions of our simulated Coma cluster (see \S\ref{sec:sim}), $T_7=9.5$ and $n_{i,-3}=3.4$, so $\lambda_{ii}\approx 4.4$ kpc. This is less than the mean free path derived from marginal stability above, implying Coulomb collisions sufficiently isotropize the pressure to avoid instabilities, at least in the cluster center. But this is still well above the AU scale gyroradii, implying we are safely in the collisionless damping regime.

The fact that $L_{\rm d} \sim L_{\rm MHD}$ when the plasma is marginally stable to microinstabilities raises potentially serious issues. It could imply that there is no MHD inertial range, since the Reynolds number at the Alfv\'en scale is of order unity, and parallel trans-Alfv\'enic motions simply dissipate. Indeed, recent analytic and numerical work finds an upper limit on shear Alfv\'en fluctuations of $\delta B_{\perp} /B_{0} \sim \beta^{-1/2}$ \citep{squire16, squire17}, above which the perturbation is rapidly quenched by the firehose instability. While CR-streaming driven turbulence lies below this limit, the background turbulence does not, and it is unclear whether equation \eqref{eq:minAngle2} (which relies on canonical Goldreich-Sridhar theory) still applies. Such issues lie well beyond the scope of this paper, but they raise important caveats to keep in mind. 

\subsection{Magnetic Topology and Other Unmodeled Factors}
\label{sec:topology}

As will be explained in \S\ref{sec:sim}, our numerical simulations are 1D spherically symmetric. This is primarily for two reasons - first, this simplifies the physics for ease of computation. Second, we want an apples-to-apples comparison with the simulations from \cite{wiener13a}, which also employed these simplifications for the first reason. We discuss here the possible implications of these simplifications.

\begin{figure}
\includegraphics[width=0.5\textwidth]{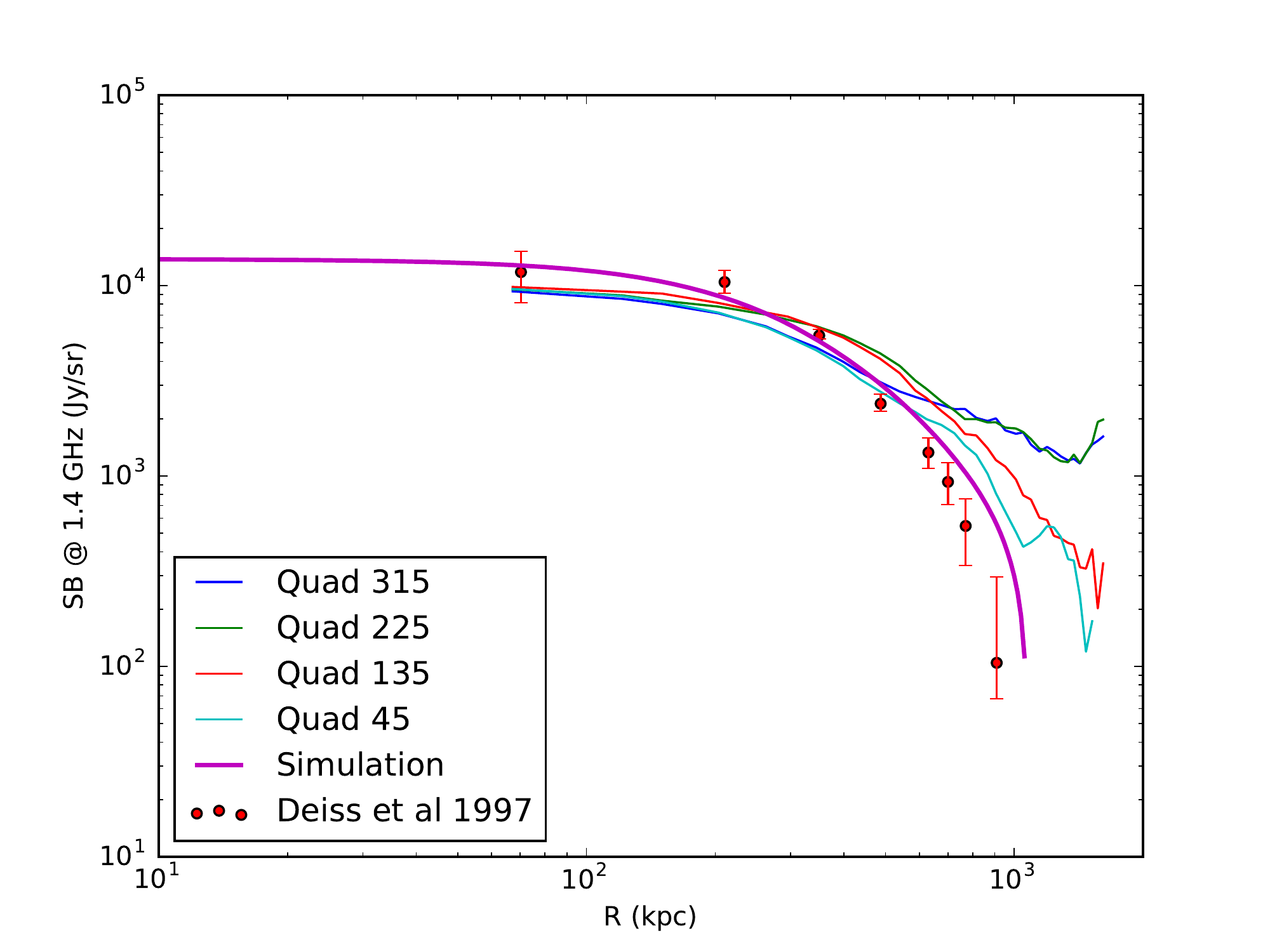}
\caption{1.4 GHz surface brightness observations of the Coma cluster. Azimuthally averaged data from \protect\cite{deiss97} (points) is compared with data from different quadrants of Coma from \protect\cite{brown11-coma} (thin lines - data provided by Larry Rudnick, personal communication). Overlayed in the thick line is the initial surface brightness of our model cluster.}\label{fig:sb}
\end{figure}

Real galaxy clusters are, of course, not spherically symmetric. The Coma cluster, which we use as our characteristic cluster, has noticeable azimuthal dependence in the radio surface brightness, as reported by \cite{brown11-coma}. We compare their 1.4 GHz observations of different quadrants of the Coma cluster using the Green Bank Telescope (provided by Larry Rudnick, personal communication) with the azimuthally averaged observations of \cite{deiss97} as well as the initial surface brightness in our model cluster in figure \ref{fig:sb}. There is clearly non-symmetric structure in the radio signal. The comparison between the two sets of data is further complicated by subtraction of the foreground signal from our own Galaxy, which is handled differently in each paper. As far as they relate to the construction of our initial CR profile, these differences in the data are of little consequence - streaming times in our simulations are not sensitive to small changes in the initial CR profile.

However, the departure from spherical symmetry has other, more significant implications for the evolution of out model CRs. The most important of these is the structure of the cluster's magnetic field. Our 1D simulations necessarily have perfectly radial magnetic fields, which is a best-case scenario for streaming. In the limit of small cross-field diffusion, CRs can only stream along magnetic field lines. They can thus leave the cluster most quickly if the field lines are radial.

But in a real cluster the magnetic field may be significantly tangled. Indeed, the very basis of our streaming model is that the presence of MHD turbulence provides a wave damping mechanism. This same turbulence will also tangle the magnetic field on some length scale, increasing the escape time of CRs in our streaming model in some potentially complicated way.

Fortunately we can make reasonable estimates of how the escape time is affected by such tangling. Suppose we want to know how long it takes for a CR to travel a distance $D$ away from the center of our cluster. In the 1D symmetric case with perfectly radial magnetic field lines, this time is just
\[
t_\tr{st}\sim D/v_\tr{st},
\]
where we simplify to a case where the streaming speed $v_\tr{st}$ is roughly constant in space and time. If instead, the field is tangled on some length scale $L$, we can treat the transport of CRs as a random walk with steps of length $L$, travel at speed $v_\tr{st}$. CR transport is then effectively diffusive with diffusion coefficient
\[
\kappa\sim Lv_\tr{st}.
\]
The time for CRs to diffuse out to a distance $D$ is then
\[
t_\tr{diff}\sim \frac{D^2}{\kappa} \sim \frac{D}{L}\frac{D}{v_\tr{st}}\sim \frac{D}{L}t_\tr{st}.
\]
In other words, tangling of the field on scale $L$ increases the escape time of CRs by roughly $D/L$.

Is this a large factor? We can approximate the tangling length scale consistently by finding the scale where the kinetic energy density in turbulence equals the magnetic energy density. Below this scale, turbulent motions are too weak to bend the field lines. This occurs when
\[
\frac{1}{2}\rho v_\tr{t}^2 = \frac{B^2}{8\pi} = \frac{1}{2}\rho v_\tr{A}^2
\]
Using our definition of $L_\tr{MHD}$ as the length scale where the turbulent speed $v_\tr{t}$ is equal to the Alfv\'en speed $v_\tr{A}$ (see \S\ref{subsec:collisionality}), we see that the tangling length scale described above is just of order $L_\tr{MHD}$. Our fiducial value is 100 kpc, and our simulated cluster has radius 1 Mpc. We may then expect the effective escape speed to exceed the streaming times in our simulation by about a factor of 10.

There are also some observational constraints on the magnetic field structure in galaxy clusters which rely on Faraday rotation measure (RM) observations of radio sources in or behind the clusters. The theoretical background for this technique is described in detail by many authors (see \cite{murgia04} and \cite{feretti12} for just two examples). In this framework, the structure of the magnetic field is described by a simple power law in Fourier space, $|B_k|^2\propto k^{-n}$, between two length scales $\Lambda_\tr{min}$ and $\Lambda_\tr{max}$.

\cite{bonafede10} use RM observations in this way to constrain the magnetic field of the Coma cluster in particular. They claim that the field which best fits the RM data is tangled on scales ranging from $\Lambda_\tr{min}\sim 2$ kpc to $\Lambda_\tr{max}\sim 34$ kpc, with a steep Kolmogorov-like spectrum of $n=11/3$. This suggests that most of the power is on large scales, i.e. most of the ``steps'' in the field line random walk are of length $\Lambda_\tr{max}\sim 34$ kpc. This is reasonably close to our above estimate of $L_\tr{MHD}$. We note, however, that in general the magnetic field structure of galaxy clusters is not well constrained by observations. So while it is hard to say with any certainty, we may reasonably expect our simplified model to underestimate streaming times by a factor of 10 - 30.

In the context of our question about radio halo turnoff times, this is not a cripplingly large factor. It extends the turnoff time from hundreds of Myr to a few Gyr, still in the range of reasonable turnoff times for the hadronic model. More importantly for the context of this work, this factor is independent of the high beta effects considered here. If our older simulations from \cite{wiener13a} are off by a factor of 10, then the new simulations presented here are also off by this same factor, and our main point is unchanged. 

Of course there are non-linear effects which complicate this - as the CR density drops the streaming speed increases. If escape speeds are initially slower by a factor of 10, the overall shutoff time may be changed by an entirely different factor. Determining the effects of this non-linearity would require non-1D simulations which are beyond the scope of this work. However, we are encouraged that in the apples-to-apples comparison we make here, the high-beta effect causes an increase in the initial streaming speeds by a factor of $\beta^{1/2}\sim 8$ in the central regions of the cluster. So differences brought about by these high beta effects should at least be comparable to the effects of field tangling.

Still, it is difficult to assess the effects of field tangling and non-linear evolution other than the above speculation. If we have overestimated the field tangling scale by a factor of about 10, which is possible considering the spectrum in \cite{bonafede10} extends down to 2 kpc, then our escape speed becomes of the order of the age of the universe even without accounting for the non-linearity of the evolution. Future observational constraints on magnetic field structure in galaxy clusters may therefore prove critical to the question of long range CR streaming.

There is another point to be made about the magnetic field topology which may be important. We have talked about the effects of a tangled field, but what if the field lines never leave the cluster? Consider a single field line which extends out to some radius $R$ in the cluster before folding back on itself and returning towards the center. Then along this field line, streaming will even out the CR density out to $R$, but CRs will be unable to leak out past $R$. If all the field lines worked this way, no CRs would be able to escape, although they could spread out evenly within radius $R$. However, if some percentage of field lines leave the cluster, then CRs will always be able to leak out along these lines. The halo dropoff time would then be primarily determined by two factors - the percentage of field lines which exit the cluster, and the rate of cross field diffusion which allows CRs on trapped lines to migrate to neighboring lines which escape.

\section{Simulation Setup}\label{sec:sim}

We reproduce here the spherically symmetric \small{ZEUS} hydrodynamic simulations of the Coma cluster from \cite{wiener13a}, but with the damping rate enhanced by the factor of $\beta^{1/2}$ from the analysis in \S\ref{sec:damping}. As mentioned in \S\ref{sec:topology}, spherical symmetry is clearly a drastic simplification (L. Rudnick; personal communication) but we assume it here to keep the problem tractable and to focus on the difference between the present treatment and \cite{wiener13a}. As in \cite{wiener13a}, we model the Coma cluster with density and temperature profiles given by
\begin{equation}
\frac{n_\tr{e}}{10^{-3}\tr{cm}^{-3}}=3.4\left[1+\left(\frac{r}{294\ \tr{kpc}}\right)^2\right]^{-1.125}
\end{equation}
\begin{equation}
T=8.25\ \tr{keV}\left[1+\left(\frac{r}{460\ \tr{kpc}}\right)^2\right]^{-0.32}
\end{equation}
These profiles are taken from \cite{pinzke10}, with the density inferred from X-ray observations (\cite{briel92}).

The magnetic field is assumed to scale with gas density as
\begin{equation}
B=B_0\left(\frac{n_\tr{e}(r)}{n_e(0)}\right)^{\alpha_B},
\end{equation}
with $B_0=5\ \mu$G and $\alpha_B=0.3$, as suggested by constraints from \cite{bonafede10} (\cite{wiener13a} also considered $\alpha = 0.5$ which is more in line with \cite{bonafede10}; the choice of $\alpha = 0.3$ here is to properly compare with the results from \cite{wiener13a}. It is conservative in the sense that it implies a lower $\beta$ and therefore a weaker Landau damping effect). The above density, temperature, and magnetic field profiles are \emph{fixed} for these simulations - only the CR distribution is evolved in time.

The initial CR distribution is of the form used in \cite{pinzke10}, motivated by cosmological hydrodynamic simulations of galaxy clusters where cosmic rays are accelerated via diffusive shock acceleration:
\begin{equation}
f_\tr{p}(r,p_\tr{p})=C(r)\sum\limits_i\Delta_i p_\tr{p}^{-\alpha_i}
\end{equation}
\begin{equation}\label{initialdist}
\mathbf{\Delta}=(0.767,0.143,0.0975)\qquad\mathbf{\alpha}=(2.55,2.3,2.15).
\end{equation}
Here and in the equations to follow, the momenta $p$ are expressed in units of $mc$ for the appropriate $m$. Namely, for actual momentum $P$ we will work in terms of $p_p = P / m_pc$ and $p_e = P / m_ec$. In this framework we have to be careful how we define our distribution functions $f_p$ and $f_e$. To be unambiguous, let us define them as such:
\[
\tr{d}n_p(r, p_p) = f_p(r, p_p)\tr{d}p_p
\]
\[
\tr{d}n_e(r, p_e) = f_e(r, p_e)\tr{d}p_e
\]
where d$n_p(r, p_p)$ is the differential number density of CRp at radius $r$ and unitless momentum $p_p$, and similarly for the electrons.

We choose the normalization $C$ to be of the form \footnote{Although this resembles the normalization in \cite{pinzke10}, please note the difference between our $C(r)$ (which have dimensions of number density) used here and the $\tilde{C}(r)=C(r)\rho(r)/m_p$ (unitless) used by \cite{pinzke10}. We utilize this formula as a convenience, but our CR density profile is actually much flatter than in their simulation.}
\begin{equation}
C(r)= \frac{(C_{\rm vir}-C_{\rm center})}{1 + \left( \frac{r}{r_{\rm trans}} \right)^{-\beta_{\rm C}}} + C_{\rm center}.
\end{equation}
In \cite{pinzke10}, the parameters $C_\tr{vir}$, $C_\tr{center}$, $r_\tr{trans}$, and $\beta_\tr{C}$ are determined from scaling relations. We instead choose values that roughly reproduce the observed synchrotron radiation. For Coma, these are $C_{\rm center}=6 \times 10^{-11} \, {\rm cm^{-3}}$, $C_{\rm vir} = 5.2 \times 10^{-11} \, {\rm cm^{-3}}$, $r_{\rm trans}=55$ kpc, $\beta_{\rm C}=1.09$. 

With the above initial distribution we evolve the CRp distribution function forward in time according to
\begin{equation}
\label{eq:fokkerplanck}
\begin{split}
\pp{f_p}{t}+(\mathbf{u}+\mathbf{v_\tr{A}})\cdot\nabla f_p= \frac{1}{3}p\pp{f_p}{p}\nabla\cdot(\mathbf{u}+\mathbf{v_\tr{A}}) \\ +\frac{1}{p^3}\nabla\cdot\left(\frac{\Gamma_\tr{damp}B^2\mathbf{n}}{4\pi^3m_p\Omega_0v_\tr{A}}\frac{\mathbf{n}\cdot\nabla f_p}{|\mathbf{n}\cdot\nabla f_p|}\right)
\end{split}
\end{equation}
as in \cite{skilling71}. In the above, $\mathbf{u}$ and $\mathbf{v_\tr{A}}$ are the gas and Alfv\'en velocities, $\Omega_0$ is the non-relativistic gyrofrequency, and $\mathbf{n}$ is a unit vector which points along the magnetic field. The last term represents diffusion with respect to the Alfv\'en wave frame due to wave damping, and is directly affected by the discussion in \S\ref{sec:damping}. More details of this equation and its numerical evolution can be found in \cite{wiener13a}. We set $\mathbf{u}\equiv 0$ in the simulations discussed here, but include it in eqn. (\ref{eq:fokkerplanck}) for completeness.

As the CRp distribution function evolves, we derive from it at every time step a steady-state secondary CRe distribution function according to
\begin{equation}
f_e(r, p_e)=\frac{1}{|\dot{p_e}|}\int_{p_e}^\infty \tr{d}p_e' s_e(r, p_e'),
\end{equation}
which balances the source function from the CRp hadronic collisions
\begin{equation}
\label{eq:source}
s_e(r, p_e)=\frac{4}{3}\frac{16m_e}{m_p}cn_N(r)\sigma_\tr{pp}f_p\left(r,p_p=\frac{16m_e}{m_p}p_e\right)
\end{equation}
with the losses from synchrotron emission and inverse Compton (IC) scattering
\begin{equation}
\dot{p_e}(r, p_e)=\frac{4}{3}\frac{\sigma_\tr{T}p_e^2}{m_ec}(\varepsilon_B(r)+\varepsilon_\tr{cmb}). 
\end{equation}
In (\ref{eq:source}), $n_N$ is the number density of target nucleons and we assume we are far above the pion production threshold.

This steady state model includes two implicit assumptions. First, the energy losses of the secondary CRe are dominated by synchrotron and IC losses. Namely this means we assume the CRe don't stream on time scales faster than their loss times ($\lsim$ 100 Myr), a condition we will have to check \emph{a posteriori}. This assumption will turn out to be well satisfied. Second, we assume that the source function $s_e$ is not changing significantly on these same time scales. This will turn out not to be very well satisfied, as we discuss below in \S\ref{sec:results}.

Once we have the secondary CRe distribution function at each time step, we can then determine the synchrotron emissivity at frequency $\nu$. From \cite{rybicki79},
\begin{equation}
j_\nu(r)=0.333\frac{\sqrt{3}}{2\pi}\frac{e^3B(r)}{m_ec^2}\int_0^\infty\tr{d}p_ef_e(r,p_e)F\left(\frac{\nu}{\nu_c}\right).
\end{equation}
In the above, $\nu_c=3eBp_e^2/4\pi m_ec$ and the function $F$ is an integral of a modified Bessel function, $F(x)=x\int_x^\infty K_{5/3}(x')\tr{d}x'$. With the emissivity in hand, we determine surface brightness and total luminosity from simple spatial integrals\footnote{We reiterate that the hadronic model for giant radio haloes is disfavored by gamma-ray non-detections. We include a radio surface brightness prediction for direct comparison with \cite{wiener13a}. The dimming of the model radio halo can also be thought of as a simple proxy for the evolution of CRp.}.

\section{Results}\label{sec:results}

\begin{figure}
\includegraphics[width=0.5\textwidth]{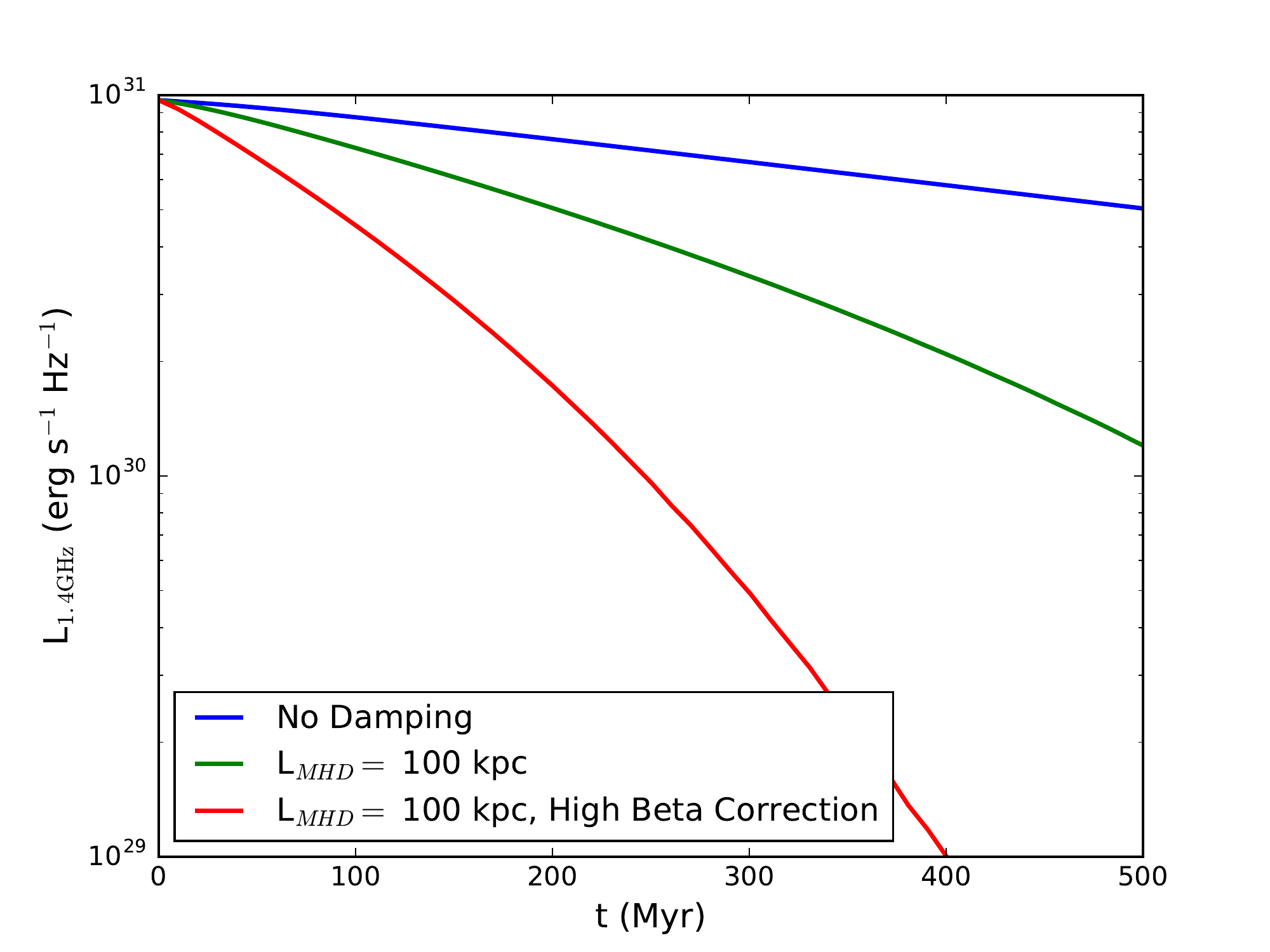}
\caption{Total 1.4 GHz luminosity as a function of time for the Coma cluster simulation. The quicker streaming speeds in the new simulation result in a quicker turn-off of the rxhibited radio halo.}\label{fig:luminosity}
\end{figure}

\begin{figure}
\includegraphics[width=0.5\textwidth]{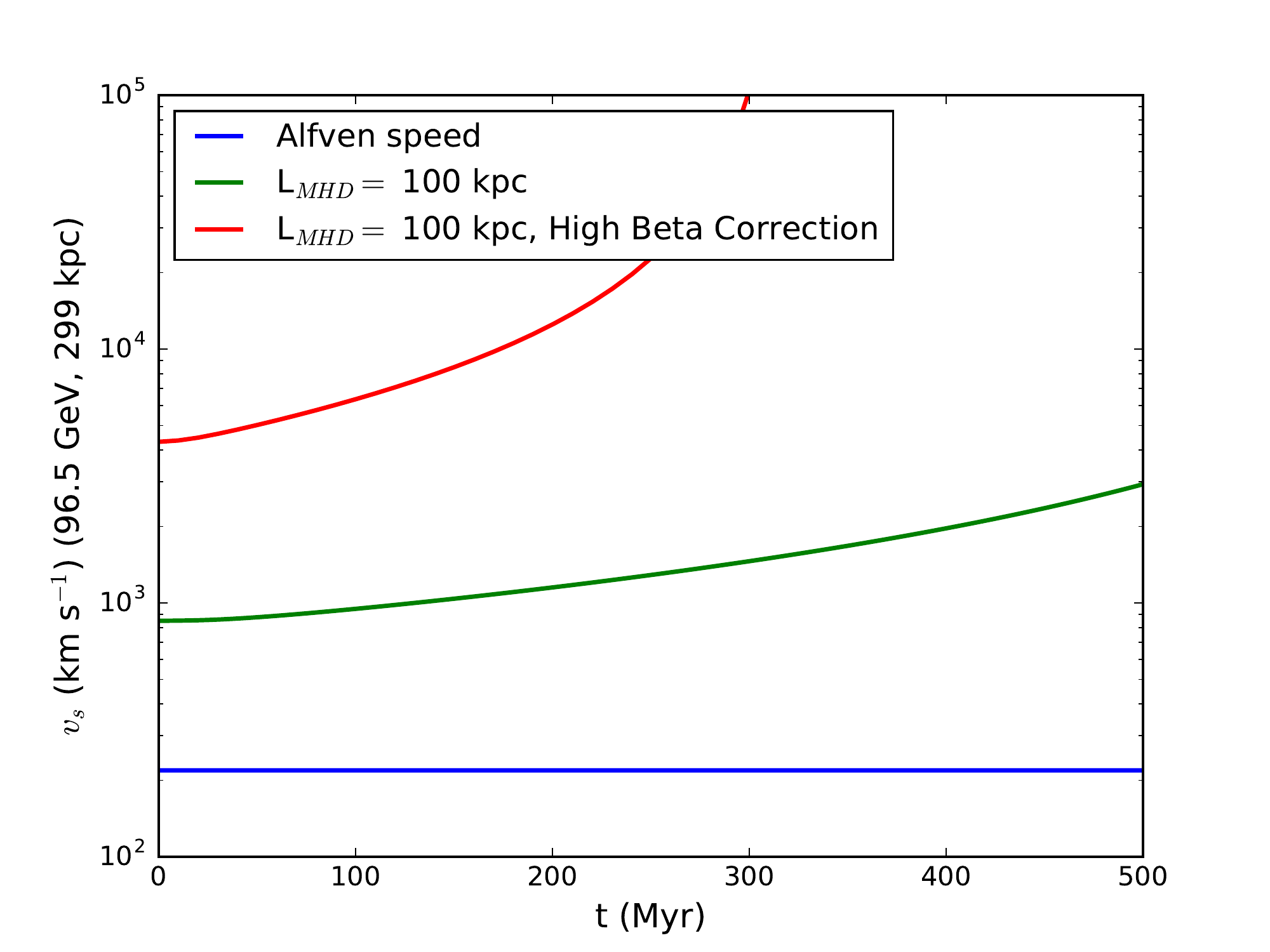}
\caption{Streaming speeds as a function of time for 100 GeV CR protons at a radius of 300 kpc in the Coma cluster simulation. The enhanced damping due to high $\beta$ effects significantly increases the streaming speed.}\label{fig:vstr100}
\end{figure}

The total 1.4 GHz luminosity of our simulated Coma cluster as a function of time is shown in Figure \ref{fig:luminosity}. Three CR transport models are compared: in the first, there is no wave damping and CRs stream at the Alfv\'en speed. In the second, there is wave damping according to \eqref{Gammaturb}. In the third, there is wave damping with the high-$\beta$ correction factor, \eqref{eq:betadamp}. We see that the increased factor of $\beta^{1/2}$ in the damping rate allows CRs to stream out even faster than in our original simulation, causing the radio luminosity to also drop on faster time scales. The streaming speeds for the relevant 100 GeV CRp are shown in figure \ref{fig:vstr100}. This figure shows the streaming speed at a fixed radius of 300 kpc as a function of time. Initially the streaming speeds are larger with the high-$\beta$ correction in comparison to without, and this enhancement grows with time due to the non-linear evolution of the streaming speeds.

We may wonder if these enhanced streaming speeds interfere with any of our assumptions, namely our assumption that CRe secondary losses are dominated by synchrotron and inverse Compton losses. CRe will stream at the same speeds as CRp of the same energy, and if the CRe stream out on time scales comparable to their loss times, our steady state treatment is incorrect. Looking at Figure \ref{fig:vstr100} we may suspect that this is the case, as the $\sim 10^4$ km/s streaming speeds imply streaming times of $\sim$100 Myr across our 1-2 Mpc cluster.

\begin{figure}
\includegraphics[width=0.5\textwidth]{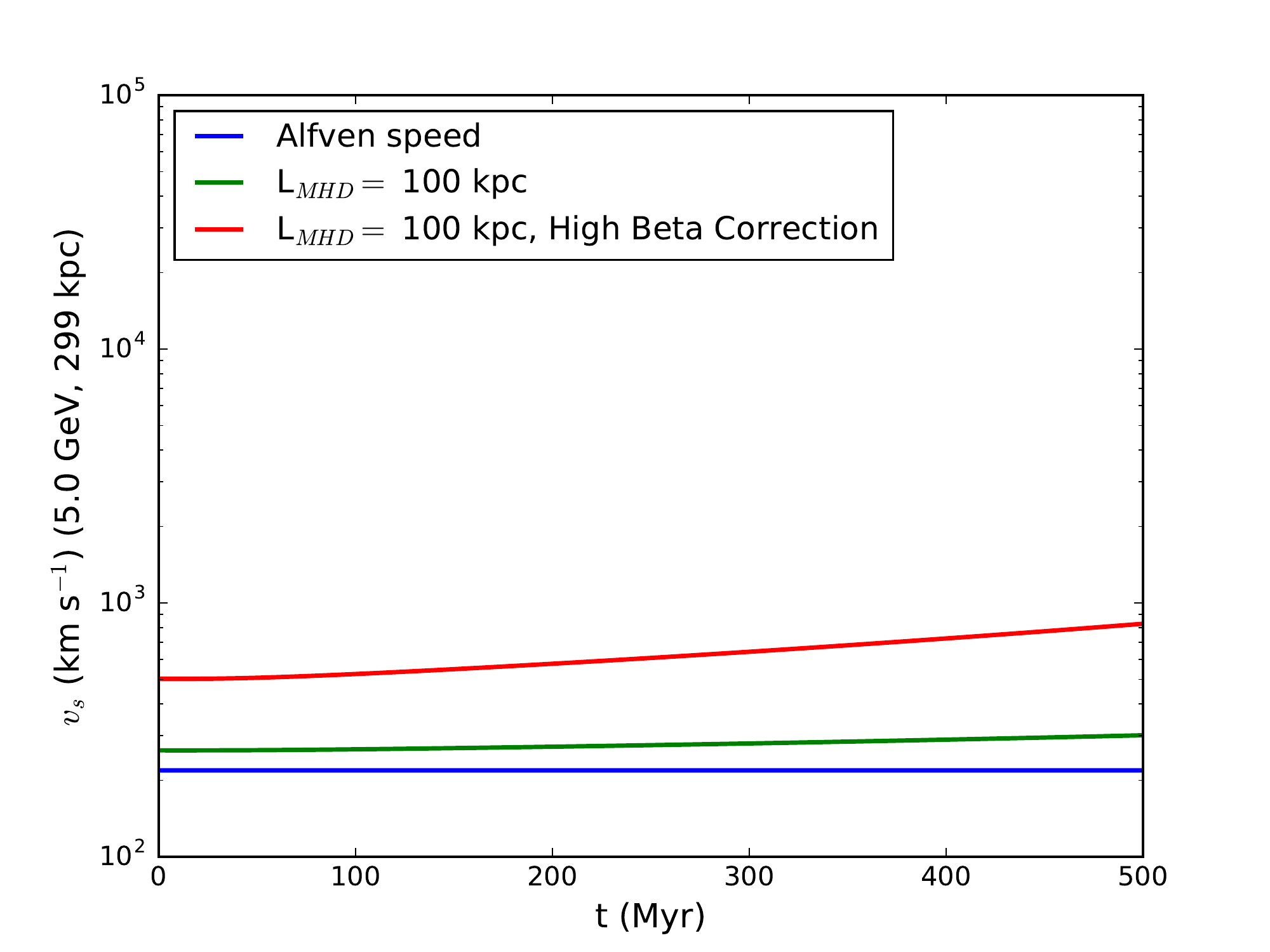}
\caption{Streaming speeds as a function of time for 5 GeV CR protons at a radius of 300 kpc in the Coma cluster simulation. At these lower energies, the streaming speeds are not significantly super-Alfv\'enic.}\label{fig:vstr5}
\end{figure}

However, for the electrons, the relevant energies for 1.4 GHz emission are 1-10 GeV. As shown in Figure \ref{fig:vstr5}, CRp at these energies stream out much more slowly, even in our high-$\beta$ model\footnote{Note that most CR energy resides at $\sim$GeV energies, and that even with the high $\beta$ correction these CRs remain more or less locked to the wave frame. Thus, CR heating in cluster cores \citep{loewenstein91,guo08a,pfrommer13,ruszkowski17} remains viable.}. We can expect the streaming times of our CRe secondaries to be $\sim$1-2 Gyr, much longer than their loss times. Note that since the streaming instability growth rate is proportional to $n_{\rm CR}$ (equation \eqref{Gammacr}), and $n_{\rm CR,p} \gg n_{\rm CR,e}$, the CRe actually scatter off the wave field created by the much more abundant CRp. If they manage to achieve spatial separation, the CRe could in principle stream much faster.

Still, the streaming speeds of the 100 GeV CRp which source these electrons are high, implying a short crossing time. Our steady state model assumes that the CRp distribution changes on time scales much longer than the CRe loss times, such that the production of secondaries `quickly' reaches an equilibrium before the CRp density changes too much. This assumption is no longer satisfied, so a more complete treatment of secondary CRe production and transport may be necessary. However, we have used somewhat artificial initial conditions which do not reflect the cosmological build up of CRp. As long as the injection time is longer than the streaming time (likely to be true for CRs sourced by structure formation (giant radio halos), less so for CRs sourced by AGN activity (radio mini-halos)), CRs will have a flat distribution, which slowly increases in normalization.

\section{Conclusion}\label{sec:conclusion}
Despite significant advances in observations and modeling, the origin and transport of cosmic rays in galaxy clusters remains incompletely understood. As long as this is the case, radiative diagnostics based on cosmic rays will be provisional, and a full assessment of the role of cosmic rays in the dynamics and energy balance of clusters will elude us.

In this paper, we have added a new ingredient to the propagation problem: the role of ion Landau damping in suppressing the growth of Alfv\'en waves excited by cosmic ray streaming.  Although waves which propagate exactly along the background magnetic field $\mbfB_0$ are undamped, inhomogeneity in $\mbfB_0$ precludes perfectly parallel propagation. At the large $\beta\equiv 8\pi P_g/B^2$ values characteristic of galaxy clusters, oblique waves are subject to strong Landau damping due to their parallel electric fields.

When the propagation angle is calculated assuming $\mbfB_0$ is a uniform field with a superimposed anisotropic MHD turbulent cascade, the resulting damping rate \eqref{eq:betadamp} is of the same form as the damping rate estimated from the turbulent shearing rate \eqref{Gammaturb}, which was taken as the dominant damping mechanism in earlier work \citep{wiener13a}, but exceeds the turbulent damping rate by a factor of $\beta^{1/2}$. It is worth noting that the assumption of an anisotropic turbulent cascade is key to this result: if the scale of variation of $\mbfB_0$ were taken to be a global scale, the resulting damping rate would be very slow \citep{zweibel03}.

Since Landau damping is a collisionless process, it is important to check whether the effective ion mean free path is much longer than the wavelengths of the Alfven waves. Estimating the mean free path $\lambda_{ii}$ to Coulomb scattering is straightforward and the result \eqref{eq:coulomb_mfp} shows that the waves are indeed collisionless with respect to Coulomb interactions. However, we also estimated for the first time the mean free path $\lambda_i$ due to scattering from microscale instabilities driven by pressure anisotropy (\ref{eq:marginal_mfp}), and found  that it too is much larger than the wavelengths of the Alfv\'en waves in question, fully  validating the role of Landau damping in suppressing the Alfv\'en waves that scatter cosmic rays in turbulent galaxy cluster plasmas. Note that there are potentially  important caveats about the nature of MHD turbulence in a high $\beta$ plasma \citep{squire16, squire17}, which are beyond the scope of this paper.

We then reconsidered the evolution of an idealized version of the radio halo of the Coma cluster following \cite{wiener13a}, but including Landau damping. In this model, the halo is produced by secondary electrons and positrons created in hadronic interactions between cosmic ray protons and thermal cluster gas. As expected, transport of the cosmic ray protons is much faster, quickly depleting them and turning off the halo on even faster timescales than predicted in the original model. This strengthens the case for rapid evolution of radio halos, at least in this idealized (1D, radial magnetic field) model cluster, unless the cosmic ray primary proton source is continuously replenished. This setup is the most optimistic case there is for evolution via streaming, an important caveat of this work. Future simulations with higher dimensionality and more complicated field topologies are necessary to study this effect in real clusters.
 
The problem of cosmic ray electron transport on galaxy clusters was first raised by \cite{jaffe77}, who pointed out that the radiative loss time of cosmic ray electrons is much shorter than their transport time at the Alfv\'en speed from the core of the cluster. If the primary CRe could stream out to large distances in less than one cooling time, they could supply the radio emission in the outskirts without being reaccelerated. However, even in our ideal scenario this does not seem to be the case. Although we have not undertaken a detailed comparison of Jaffe's model with ours, the transport speeds of 5 GeV electrons shown in Figure \ref{fig:vstr5} are well below the $\sim 2000$ km/s transport speed that Jaffe estimated was necessary to form the halo with primary electrons.
 
The rapid transport speed of cosmic ray protons due to Landau damping of Alfv\'en waves could help to explain the stringent upper limits on diffuse $\gamma$-ray emission from galaxy cluster cores reported by \cite{pinzke10,  ensslin11, ahnen16}. It also has important consequences for hadronic production of CRe `seeds' for turbulent reacceleration - rapid streaming produces a flat CRp profile, giving a seed population which better fits observational constraints \citep{pinzke17}. It is also relevant to the tension between observed radio halo luminosities and models of CR heating of cluster cores \citep{jacob17}, since high energy CRs responsible for the former stream at much higher velocities than the lower energy CRs relevant for the latter. These issues are beyond the scope of the current paper, but a topic for future work.

\noindent{Acknowledgements:}

We are happy to acknowledge discussions with Matt Kunz, Eliot Quataert, and Larry Rudnick. JW and EGZ acknowledge support by NSF Grant AST-1616037, the WARF Foundation, and the Vilas Trust. SPO acknowledges support from NASA grant NNX15AK81G.
\bibliography{master_references2}

\end{document}